# Exponentially Stabilizing Continuous-Time Controllers for multi-domain hybrid systems with application to 3D bipdeal walking

Chunbiao Gan, Haihui Yuan, Shixi Yang and Yimin Ge

*Abstract*—This paper presents a systematic approach to exponentially stabilize the periodic orbits of multi-domain hybrid systems arising from 3D bipedal walking. Firstly, the method of Poincaré sections is extended to the hybrid systems with multiple domains. Then, based on the properties of the Poincaré maps, a continuous piecewise feedback control strategy is presented, and three methods are furthermore given to design the controller parameters based on the developed theorems. By those design methods, the controller parameters in each continuous phase can be designed independently, which allows the strategy to be applied to hybrid systems with multiple domains. Finally, the proposed strategy is illustrated by a simulation example. To show that the proposed strategy is not limited to bipedal robots with left-right symmetry property which is assumed in some previous works, an underactuated 3D bipedal robot with asymmetric walking gait is considered.

## I. Introduction

This paper addresses the problem of exponentially stabilizing the periodic orbits of multi-domain hybrid systems arising from 3D bipedal walking [1-5]. Although there is a considerable body of work on the exponentially stable walking control for planar bipeds [6-10], the exponentially stable walking of the 3D bipeds is still far from being well solved. One of the key features of the problem is that, compared with the planar biped, the 3D bipedal walking system is not only hybrid [6, 9-11], but of multiple domains (or continuous phases) [1, 12].

For a planar biped, a common "trick" in the field is to develop a model of the biped with one of the legs in contact with the ground, and then to "swap" or relabel the joint angles after the swing right leg impact with the ground in order to propagate the dynamics for the ensuing step [1]. In this way, the number of the domains can be reduced by half. Thus, despite the fact that the dynamical model of a planar biped has two continuous phases, the planar walking is often modeled with only a single domain. However, when it comes to the dynamical model of a 3D biped, one must "flip" the sign on the hip width and "flip" the sign convention for all joint angles that are not in the sagittal plane [1]. As a result, the dynamical models in the left and right support phase are not absolutely symmetric. In other words, the number of the domains in the dynamical model can't be reduced. Therefore, the 3D bipeds are naturally modeled as multi-domain hybrid systems. However, to the best of our knowledge, few results have been focused on exponentially stabilizing the periodic orbits of multi-domain hybrid systems.

In the early work [8], the optimal control method is successfully applied to achieve the exponetial stability of a planar running gait with two continuous phases. However, this method requires a high computational cost since the optimization procedure designs the gait and the controller simultaneously [8, 13]. During the optimization procedure, all the feedback control parameters within each continuous phase as well as the virtual constraint parameters, i.e., the gait design parameters, have to be taken as optimization variables. In [8], the running gait of a five-link planar biped with two continuous phases has 23 independent optimization variables. Compared with the planar bipeds, the 3D ones have more degrees of freedom and more complicated dynamical equations. Considering that the amount of computational cost is exponentially proportional to the number of optimization variables, it would result in a huge computational cost. Therefore, technical limits exist in extending this optimal control method to the 3D bipedal walking with multiple continuous phases.

In the later works [5, 14, 15], the well-known event-based feedback control method is successfully applied to achieve the exponential stability of a 3D bipedal walking gait with two continuous phases. In these studies, two partial maps (or one-step maps [15]) are firstly constructed corresponding to the two continuous phases. Then based on the left-right symmetry property of the walking gait, one of the two partial maps is transformed into a linear relationship while the event-based feedback control is imposed on the other partial map. In this way, the two-domain system can be dealt with just like a single-domain system. In [16], the event-based feedback control method is further developed into a time-invariant hybrid control strategy on the basis of symmetry. However, this strategy is performed mainly for symmetric walking. Moreover, the work does not give a clue on how to extend the strategy into the hybrid systems with multiple domains.

Recently, A BMI optimization method is applied to exponentially stabilizing the periodic orbits of hybrid systems [17]. Firstly, a parameterized family of continuous time controller is designed for the hybrid system. By assuming that the periodic orbit is invariant under the choice of controller parameters, the problem of exponential stabilization was translated into a set of Bilinear Matrix Inequalities (BMIs). Then, a BMI optimization problem is set up to tune the parameters of the continuous-time controllers so that the Jacobian of the Poincaré map has its eigenvalues in the unit circle. By introducing continuous-time feedback laws in each continuous phase of a hybrid system, this method can be applicable even for a hybrid system with multiple domains. Unfortunately, during this method, the controller parameters in each continuous phase need to be designed simultaneously, which would become challenging when the number of continuous phases significantly increases. In addtion, the assumption of invariant periodic orbits for the closed-loop hybrid system would be a potential limitation of this method [17].

In this work, to address the problem of exponentially stabilizing the periodic orbits of multi-domain hybrid systems arising from 3D bipedal walking, we develop a continuous piecewise feedback control strategy. In section II, the method of Poincaré sections is extended to the hybrid systems with multiple domains. By exploiting the properties of the Poincaré maps, a continuous piecewise feedback control strategy is presented in secion III, and three methods are furthermore given to design the controller parameters based on the design theorems developed in this paper. By those design methods, the controller parameters in each continuous phase can be designed independently. In this way, it takes almost no additional computational cost even when the number of continuous phases significantly increases. In section IV, the proposed strategy is illustrated by a simulation example. To show that the proposed strategy is not limited to bipedal robots with left-right symmetry property which is assumed in some previous works [5, 16, 17], an underactuated 3D bipedal robot with asymmetric walking gait is considered. In the final section, the results of this work are summarized and discussed.

## II. Technical background

As stated above, the 3D bipeds are naturally modeled as multi-domain hybrid systems. In addtion, it often occurs that there are three or more continuous phases in bipedal walking gaits, for example, bipedal walking with several continuous subphases [18] and bipedal running with nontrival feet. In those cases, bipedal robots would surely be modeled as multi-domain hybrid systems. In this section, we will review the hybrid models for multi-domain hybrid systems as well as the primary analysis tool that is used in this paper: the Poincare's method.

### A. Multi-domain hybrid models

Hybrid systems with a single-domain in bipedal locomotion have been studied extensively [9-11]. Here, we are going to address systems with $N \geq 2$ continuous domains and discrete transitions between the domains. Without loss of generality, suppose that phases are executed in a fixed order $1 \to 2 \to \ldots N \to 1$, as shown in Fig. 1. Considering the cyclic behavior in bipedal locomotion, we will use the notation $N+1=1$. For compactness of notation, the multi-domain model will be denoted as a 5-tuple $\Sigma = (X, U, FG, S, \Delta)$

$$\Sigma = \begin{cases} X = \{X_i\}_{i=1}^{N} : X_i \subset R^{m_i} \\ U = \{U_i\}_{i=1}^{N} : U_i \subset R^{n_i} \\ FG = \{f_i, g_i\}_{i=1}^{N} : \dot{x}_i = f_i(x_i) + g_i(x_i)u_i , \\ S = \{S_i^{i+1}\}_{i=1}^{N} : \{x_i \mid H_i^{i+1} = 0, \dot{H}_i^{i+1} \neq 0\} \\ \Delta = \{\Delta_i^{i+1}\}_{i=1}^{N} : x_{i+1}^{+} = \Delta_i^{i+1}(x_i^{-}) \end{cases} \quad (1)$$

where $X_i$ is an open connected subset of $R^{m_i}$, upon which the autonomous differential equation $\dot{x}_i = f_i(x_i) + g_i(x_i)u_i$ of phase-$i$ is defined, $u_i \in U_i \subset R^{n_i}$ is the vector of actuator torques, $S_i^{i+1}$ is an embedded submanifold of co-dimension one in the state space $X_i$ that determines when a transition from $X_i$ to $X_{i+1}$ takes place according to the reset map $\Delta_i^{i+1}$, $x_i^-$ is final state of phase-$i$, $x_{i+1}^+$ is the initial state of phase-$i+1$.

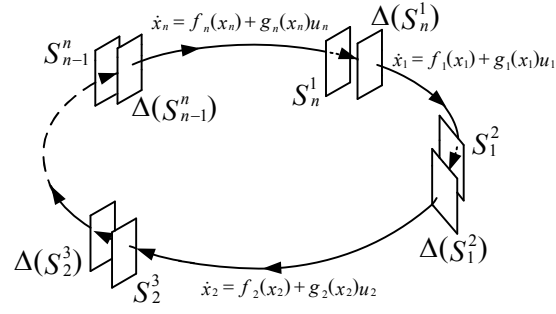

Fig. 1. The multi-domain hybrid model in periodic walking motion. $S_i^{i+1}$ $(i=1,...,n-1)$ are the switching hyper-surfaces, $\Delta S_i^{i+1}$ $(i=1,...,n-1)$ denote the hyper-surfaces after each impact event or the initial hyper-surfaces of the next domain after a nonsmooth process.

Under assumptions analogous to those for the single-domain model [9-11], a unique, maximal solution of the multi-domain model can be constructed by piecing together trajectories of the flows $FG_i$ in such a way that a transition occurs when a flow intersects a switching hyper-surface, $S_i^{i+1}$, and at each transition, the new initial condition is determined by the reset maps $\Delta_i^{i+1}$. To avoid chattering, it is assumed that a solution through a domain must have a non-zero duration [13].

### B. Periodic Orbits and the Poincare's Method

Let $X = X_1 \cup X_2 ... \cup X_N$, A solution $\varphi(t)$ of (1) is *periodic* if there exists a finite $T$ such that $\varphi(t+T) = \varphi(t)$ for all $t \in [0,\infty)$. A set $O \in X$ is a *periodic orbit* of (1) if $O \in \{\varphi(t) | t \geq t_0\}$ for for some periodic solution $\varphi(t)$. A periodic orbit $O$ is *transversal* to $S_i^{i+1}$ if its closure intersects $S_i^{i+1}$ in exactly one point, and for $x_i^* = O \cap S_i^{i+1}$, $\dot{H}_i^{i+1}(x_i^*) \neq 0$. A periodic orbit $O$ is *transversal* if it is transversal to $S_i^{i+1}$ for all *i*. In the case of a bipedal robot, a nontrivial, transversal, periodic orbit will also be referred to as *periodic locomotion* [13].

The definitions of orbital stability in the sense of Lyapunov, orbital asymptotic stability, and orbital exponential stability are analogous to those systems with single-domain. The Poincaré return map remains the mathematical tool for determining the existence and stability properties of periodic orbits. Define the phase-*i* *time-to-impact function*, $T_{I,i}: x_i \to R \cup \infty$, by

$$T_{I,i}(x_0) := \begin{cases} \inf\{t \geq 0 \mid \phi_i(t, x_0) \in S_i^{i+1}\} & \text{if } \exists t \text{ such that } \phi_i(t, x_0) \in S_i^{i+1} \\ \infty & \text{otherwise,} \end{cases} \quad (2)$$

where $\phi_i(t, x_0)$ is an integral curve of the differential equation $\dot{x}_i = f_i(x_i) + g_i(x_i)u_i$ defined in (1) corresponding to $\phi_i(0, x_0) = x_0$. From [13], $T_{I,i}(x_0)$ is continuous at points $x_0$ where $0 < T_{I,i}(x_0) < \infty$ and the intersection with $S_i^{i+1}$ is transversal. Hence, $\tilde{X}_i := \{x_i \mid 0 < T_{I,i}(x_0) < \infty \text{ and } \dot{H}_i^{i+1}(\phi_i(T_{I,i}(x_i), x_i)) \neq 0\}$ is open, and consequently, $\tilde{S}_{i-1}^i := \Delta_{i-1}^{i}{}^{-1}(\tilde{X}_i)$ is an open subset of $S_{i-1}^i$. Under the hypotheses in [13], the *generalized* Poincaré *phase-i map* $P_i: \tilde{S}_{i-1}^i \to S_i^{i+1}$ can be defined as a partial map

$$P_i(x_{i-1}^S) := \phi_i(T_{I,i}(\Delta_{i-1}^i(x_{i-1}^S)), \Delta_{i-1}^i(x_{i-1}^S)), \quad (3)$$

where $x_{i-1}^S$ is a state defined in $\tilde{S}_{i-1}^i$, and $P_i$ is $C^1$ in a neighborhood of $x_{i-1}^S$. It is noted that, $P_i: \tilde{S}_{i-1}^i \to S_i^{i+1}$ is often expressed as $P_i: S_{i-1}^i \to S_i^{i+1}$ for simplicity. The Poincaré return map can be defined as the composition of the generalized Poincaré phase-*i* maps, starting at any point in the cycle $1 \to 2 \to \ldots N \to 1$. Without lose of generality, we start it at $i = 1$, so that

$$P = P_N \circ P_{N-1} \circ \ldots \circ P_1. \quad (4)$$

**Theorem 1(Connecting Multi-domain Models to Single-domain Models** [1, 13]**):**

Let $P$ be the Poincaré return map defined in (4) for the multi-domain model in (1). $P$ is also the Poincaré return map for the single-domain system

$$\Sigma = \begin{cases} \dot{x}_1 = f_1(x_1) + g_1(x_1)u_1 & x_1^- \notin S_2^1 \\ x_1^+ = \Pi(x_1^-) & x_1^- \in S_2^1 \end{cases}, \quad (5)$$

where $\Pi := \Delta_N^1 \circ P_N \circ \cdots \circ P_2$

This theorem is very important because it shows that results developed for single-domain models can be applied to models with multiple domains. In particular, suppose that the multi-domain hybrid model (1) is $C^1$ in each phase and has a transversal periodic orbit $O$. Then, results in [13] show that $\Pi := \Delta_N^1 \circ P_N \circ \cdots \circ P_2$ is $C^1$ in a neighborhood of $x^* = O \cap S$, and thus $P$ is $C^1$ in a neighborhood of $x^*$. Exponential stability the hybrid system can be checked by evaluating eigenvalues of the Jacobian of $P$ at $x^*$, and the system is exponential stable if the Jacobian has spectral radius smaller than 1. Moreover, according to the result in [11], the exponential stability can also be by evaluated in a reduced-dimensional state space.

**Theorem 2**: Suppose that the multi-domain hybrid model (1) is $C^1$ in each phase and has a transversal periodic orbit $O$. Let $A = \dfrac{\partial P}{\partial x_1^S}(x_1^*)$ be the Jacobian matrix of the Poincaré return map $P$. Let $A_i = \dfrac{\partial P_i}{\partial x_i^S}(x_i^*)$ be the Jacobian matrix of $P_i$. Then

$$A = A_N A_{N-1} \cdots A_1.$$

*Proof.* From Theorem 1, we can know that the Poincaré return map $P$ is $C^1$ in a neighborhood of $x^* = O \cap S$. Then, according to the chain rule for computing the derivative of the composition of two or more functions, the Jacobian $A$ for $P$ can be expressed as

$$\begin{aligned}
\frac{\partial P}{\partial x_1^S}(x_1^*) &= \frac{\partial (P_N \circ \cdots \circ P_2 \circ P_1)}{\partial x_1^S}(x_1^*) = \frac{\partial (P_N \circ \cdots \circ P_2)}{\partial (P_1(x_1^S))}(P_1(x_1^*)) \cdot \frac{\partial P_1}{\partial x_1^S}(x_1^*) \\
&= \frac{\partial (P_N \circ \cdots \circ P_2)}{\partial x_2^S}(x_2^*) \cdot \frac{\partial P_1}{\partial x_1^S}(x_1^*) = \frac{\partial (P_N \circ \cdots \circ P_3)}{\partial (P_2(x_2^S))}(P_2(x_2^*)) \cdot \frac{\partial P_2}{\partial x_2^S}(x_2^*) \cdot \frac{\partial P_1}{\partial x_1^S}(x_1^*) \\
&= \frac{\partial P_N}{\partial x_N^S}(x_N^*) \cdots \frac{\partial P_2}{\partial x_2^S}(x_2^*) \cdot \frac{\partial P_1}{\partial x_1^S}(x_1^*) \\
&= A_N A_{N-1} \cdots A_1
\end{aligned} \quad (6)$$

which completes the proof.

### III. Exponentially stabilizing the periodic orbits of multi-domain hybrid systems

As stated before, the BMI optimization method can be applied to exponentially stabilizing the periodic orbits of multi-domain hybrid systems. However, during this method, the controller parameters in each continuous phase need to be designed simultaneously, which would result in huge computational cost when the number of continuous phases significantly increases. In addtion, the assumption of invariant periodic orbits for the closed-loop hybrid system would be a potential limitation of this method. To address these two problems, in this section, a continuous piecewise feedback control strategy is proposed, and two design theorems are furthermore given to design the controller parameters in each phase independently.

#### A. The continuous piecewise feedback control strategy

For hybrid systems with periodic orbits, the design of the controller is equivalent to the design of the Jacobian of the Poincaré return map. From theorem 2, one can clearly see that the Jacobian of the Poincaré return map can be expressed as the successive multiplies of the Jacobians of partial maps. This leads us to think that the design of the Poincaré return map's Jacobian can be realized by imposing continuous feedback in each continuous phase to design the Jacobian of each partial map.

Consider the multi-domain hybrid system with periodic orbit defined in (1). For each phase-$i$, suppose that the continuous-time controller can be expressed as the following parameterized control law

$$u_{i,\beta} = \Gamma_i(x_i, \beta_i), \quad (7)$$

Where $\beta_i \in \Xi_i$ and $\Xi_i \subset R^p$ represent the finite-dimensional parameter vector in an open neighborhood of 0 and set of admissible parameters, respectively, for some positive integer $p$. Moreover $\Gamma_i(x_i, \beta_i): X_i \times \Xi_i \to U_i$ is a $C^{n_p}$ map, where $n_p$ is some positive integer larger than 1. In contrast with the BMI method in [17], where the continuous-time controller $\Gamma_i(x_i, \beta_i)$ is designed to meet $\partial \Gamma_i(x_i, \beta_i)/\partial \beta_i = 0$, so that the periodic orbit will be invariant under the choice of the control parameters $\beta_i$, in this paper, we consider $\partial \Gamma_i(x_i, \beta_i)/\partial \beta_i \neq 0$. In particular, $\Gamma_i(x_i, \beta_i)$ will be designed such that

C1) $\lim_{\beta_i \to 0} \Gamma_i(x_i, \beta_i) = u_i(x_i)$

C2) $\lim_{\beta_i \to 0} \partial \Gamma_i(x_i, \beta_i)/\partial x_i = \partial u_i(x_i)/\partial x_i$.

Conditions C1 and C2 are required so that the periodic orbit of the hybrid system will not be significantly changed by the parameter vector $\beta_i$ which is designed to be small enough, see (42) and (43). In order to have a more specific understanding about the continuous-time controller $\Gamma_i(x_i, \beta_i)$, the following lemma is presented.

**Lemma 1**. Suppose that $u_i(x_i)$ defined in (1) can be expressed by a general form $u_i(x_i) = \Upsilon_1(\Upsilon_2(x_i)) + \Upsilon_3(x_i)$, where $\Upsilon_1$, $\Upsilon_2$, and $\Upsilon_3$ are all $C^1$ functions with respect to $x_i$. If there exists a continuous function $\Omega(x_i, \beta_i)$ which is $C^{n_p}$ functions with respect to $(x_i, \beta_i)$ such that $\lim_{\beta_i \to 0} \Omega(x_i, \beta_i) = 0$ and $\lim_{\beta_i \to 0} \partial \Omega(x_i, \beta_i)/\partial x_i = 0$. Then both $\Gamma_1(x_i) = \Upsilon_1(\Upsilon_2(x_i) + \Omega(x_i, \beta_i)) + \Upsilon_3(x_i)$ and $\Gamma_2(x_i) = \Upsilon_1(\Upsilon_2(x_i)) + \Upsilon_3(x_i) + \Omega(x_i, \beta_i)$ meet the conditions C1 and C2.

*Proof*. The proof can be straightly obtained.

Since the designed controller $\Gamma_i(x_i, \beta_i)$ is $C^{n_p}$ with respect to $\beta_i$, integrating the differential equation $\dot{x}_i = f_i(x_i) + g_i(x_i)\Gamma_i(x_i, \beta_i)$, we can easily know that the integral curve of the differential equation, $\phi_{i,\beta}(t, x_0, \beta_i)$, is also $C^{n_p}$ with respect to $\beta_i$. Therefore, the generalized Poincaré phase-$i$ map $P_{i,\beta}(x_{i-1}^S, \beta_i) := \phi_i(T_{I,i}(\Delta_{i-1}^i(x_{i-1}^S)), \Delta_{i-1}^i(x_{i-1}^S), \beta_i)$ is $C^{n_p}$ with respect to $\beta_i$. Recall that $P_{i,\beta}$ is continuously differentiable with respect to $x_{i-1}^S$. Therefore, the total derivative of $P_{i,\beta}$ can be expressed as

$$dP_{i,\beta} = \frac{\partial P_{i,\beta}}{\partial x_{i-1}^S} dx_{i-1}^S + \frac{\partial P_{i,\beta}}{\partial \beta_i} d\beta_i. \tag{8}$$

In the next step, if we design a continuous feedback law such that

$$\beta_i = -K_i(x_{i-1}^S - x_{i-1}^*), \tag{9}$$

where $K_i \in R^{p \times n}$ is the feedback parameter matrix. Replacing (9) in (8) yields

$$A_i^\beta = A_{i,\beta} - F_{i,\beta} K_i, \tag{10}$$

where $A_i^\beta = dP_{i,\beta}/dx_{i-1}^S$ is the designed Jacobian of partial map $P_{i,\beta}$, $A_{i,\beta} = \partial P_{i,\beta}/\partial x_{i-1}^S$ is the partial derivative of $P_{i,\beta}$ with respect to $x_i$, and $F_{i,\beta} = \partial P_{i,\beta}/\partial \beta_i$ is the partial derivative of $P_{i,\beta}$ with respect to $\beta_i$.

Suppose that we have designed the Jacobians $A_i^d$ for all $i$ independently by following the above procedures, which will be done in the next subsection. Then, according to theorem 2, the Jacobian of the Poincare return map can be designed as

$$A^\beta = A_N^\beta A_{N-1}^\beta ... A_1^\beta. \tag{11}$$

Let $\rho(A_N^\beta A_{N-1}^\beta ... A_1^\beta)$ be the spectral radius of $A_N^\beta A_{N-1}^\beta ... A_1^\beta$. From theorem 1, the closed-loop hybrid system is exponentially stable, if $\rho(A_N^\beta A_{N-1}^\beta ... A_1^\beta) < 1$, which is the ultimate goal of this work.

*B. The picewise feedback controller*

As stated before, in some previous works, the controller parameters in each continuous phase need to be designed simultaneously, which would become challenging when the number of continuous phases significantly increases. In this subsection, we present two design theorems so that we can design the feedback parameters in each continuous phase independently.

Taking the designed controlle $\Gamma_i(x_i, \beta_i)$ and the feedback law (9) in the above subsection into account, the multi-domain hybrid model defined in (1) will be updated as 6-tuple $\sum_{cl} = (X, U, \Theta, FG, S, \Delta)$

$$\sum_{cl} = \begin{cases} X = \{X_i\}_{i=1}^{N} : X_i \subset R^{m_i} \\ U = \{U_i\}_{i=1}^{N} : U_i \subset R^{n_i} \\ \Theta = \{\Theta_i\}_{i=1}^{N} : \beta_i = -K_i(x_{i-1}^S - x_{i-1}^*) \\ FG = \{f_i, g_i\}_{i=1}^{N} : \dot{x}_i = f_i(x_i) + g_i(x_i)\Gamma_i(x_i, \beta_i) \\ S = \{S_i^{i+1}\}_{i=1}^{N} : \{x_i \mid H_i^{i+1} = 0, \dot{H}_i^{i+1} \neq 0\} \\ \Delta = \{\Delta_i^{i+1}\}_{i=1}^{N} : x_{i+1}^+ = \Delta_i^{i+1}(x_i^-) \end{cases} \quad (12)$$

where $\{\Theta_i\}_{i=1}^{N}$ is piecewise defined to impose feedbacks, for convenience, we call it the piecewise feedback controller.

According to Appendix A, when we design the Jacobian of each partial map, (10) can be expressed in a simple form

$$A_i^d = A_i - F_i K_i, \quad (13)$$

which is only dependent on the feedback parameter matrix $K_i$. Therefore, according to theorem 2, if we have designed the piecewise feedback controller $\{\Theta_i\}_{i=1}^{N}$, the Jacobian of the Poincare return map would be designed as

$$A^d = A_N^d A_{N-1}^d ... A_1^d. \quad (14)$$

Let $\rho(A_N^d A_{N-1}^d ... A_1^d)$ be the spectral radius of $A_N^d A_{N-1}^d ... A_1^d$. Then our goal is to design the piecewise feedback controller $\{\Theta_i\}_{i=1}^{N}$ so that $\rho(A_N^d A_{N-1}^d ... A_1^d) < 1$.

To satisfy the condition $\rho(A_N^d A_{N-1}^d ... A_1^d) < 1$, the feedback parameters in the piecewise feedback controller $\{\Theta_i\}_{i=1}^{N}$ can be designed simultaneously, as in [17].

However, it would result in huge computational cost when the number of continuous phases significantly increases. Hence, we hope to find ways to design the feedback parameters in each continuous phase independently. For this goal, the following design theorems are presented.

**Theorem 3**: For a *N*-domain hybrid closed-loop system $\sum_{cl} = (X, U, \Theta, FG, S, \Delta)$, if there exists a piecewise feedback controller $\{\Theta_i\}_{i=1}^N$, under which the Jacobian of each phase-*i*, $A_i^d \in R^{n \times n}$, is designed such that

H1) $A_i^d \in R^{n \times n}$ is symmetric;

H2) The spectral radius of $A_i^d$ meets $\rho(A_i^d) < 1$.

Then the closed-loop system is exponentially stable.

*Proof.* See Appendix A.

**Remark 1**: The Item 1 is very important. If the symmetric condition is not met, we will have no idea whether the spectral radius of $A_N^d...A_2^d A_1^d$ is acceptable, even when we have designed a set of Jacobians that all look perfect. Here, we will show you an interesting example. For a two-domain hybrid system, suppose that we have designed two Jacobians for the two partial maps such as, $A_1^d = \begin{bmatrix} 0.7 & 0.65 \\ 0 & 0.5 \end{bmatrix}$, $A_2^d = \begin{bmatrix} 0.7 & 0 \\ 0.65 & 0.5 \end{bmatrix}$. Both the spectral radiuses of them are 0.75, which is quite acceptable. However, the spectral radius of $A_2^d A_1^d$ is 1.0453, which indicates an unstable closed-loop system.

**Theorem 4**: For a *N*-domain hybrid closed-loop system $\sum_{cl} = (X, U, \Theta, FG, S, \Delta)$, if there exists a piecewise feedback controller $\{\Theta_i\}_{i=1}^N$, under which the Jacobian of each phase-*i*, $A_i^d \in R^{n \times n}$, satisfies $m_i = \max_{1 \leq j, k \leq n} |A_i^d(j,k)| \leq 1/n$, then the closed-loop system is exponentially stable.

*Proof.* See Appendix B.

From the above two theorems, we can know that the feedback parameter matrix $K_i$ in each continuous pahse can be designed independently to achieve the exponential stability of the hybrid system, as long as the designed Jacobian of each partial map satisfy some conditions. In the following subsection, three methods are presented to design the feedback parameter matrix $K_i$ based on the above theorems.

*B. Methods for Designing the Piecewise feedback controller*

The implement of the continuous piecewise feedback control strategy requires the design of the piewise feedback controller $\{\Theta_i\}_{i=1}^{N}$. And the essetial part is to obtain the feedback parameter matrix $K_i$ in each pahse independently. In the following, three different methods are given to obtain the feedback parameters.

**1) Symmetric matrix method**. The method is based on Theorem 3. To achieve the exponential stability of a *N*-domain hybrid system, a constant symmetric matrix $M_{sym}$ with spectral radius smaller than 1 is firstly picked. Then, according to Theorem 3, the hybrid system would be exponentilly stable, if the Jacobian of each phase-*i* is designed to be $M_{sym}$, i.e.,

$$A_i^d = M_{sym}, 1 \leq i \leq N. \tag{15}$$

According to (13), we obtain

$$M_{sym} = A_i - F_i K_i. \tag{16}$$

In general, $F_i$ is not a square matrix. Hence, the feedback parameter matrix $K_i$ from the above equation can be calculated as

$$K_i = pinv(F_i)(A_i - M_{sym}), \tag{17}$$

where $pinv(F_i)$ denotes the Moore-Penrose pseudoinverse of $F_i$.

During the symmetric matrix method, the feedback parameters can be independently and directly obtained with no optimization process, which allows the method to be used for real-time application.

**2) Scale factor method**. The Scale factor method is based on Theorem 4. Let $m_i = \max\limits_{1 \leq j,k \leq n} |A_i(j,k)|$ denotes the largest element of $|A_i|$, and let $c_i = 1/(n.m_i)$ be the corresponding constant scale factor. Then, from Theorem 4, it's obvious that the N-domain hybrid system is exponentially stable, if the Jacobian of each phase-i, $A_i^d$, is designed to be

$$A_i^d = c_i A_i, 1 \leq i \leq N. \tag{18}$$

Then, combining (13) and (18), we obtain

$$c_i A_i = A_i - F_i K_i. \tag{19}$$

Similar to (17), the feedback matrix $K_i$ can be calculated as

$$K_i = pinv(F_i)(1-c_i)A_i. \tag{20}$$

In the Scale factor method, different Jacobians are designed with different coefficients, i.e. $c_i$. Therefore, compared with the symmetric matrix method, the scale factor method accounts for the differences between the Jacobians of partial maps.

**3) DLQR-based method**. The conventional DLQR method has been extensively used to design controllers. Here, it wil be extended to the multi-domain hybrid system. Starting by defining $Y_i = x_{i-1}^S - x_{i-1}^*$, the linearization of equation (13) can be expressed as

$$Y_{i+1} = A_i Y_i + F_i \beta_i. \tag{21}$$

During the DLQR method, the feedback gain matrix $K_i$ is calculated via a discrete linear quadratic regulator (DLQR) so that the feedback control law $\beta_i = -K_i Y_i$ minimizes the cost function $\sum_k (Y'_i Q_i Y_i + \beta'_i R_i \beta_i)$ subject to the state dynamics (21), where $Q_i$ is a real positive semi-definite weighting matrix associated with the robustness of the system, and $R_i$ is a real positive definite weighting matrix associated with the input energy, i.e., $\beta_i$. The two weighting matrices are often decided using trial and error method. For simplicity, they are often designed to be a unit matrix multiplied by a positive constant number. Now that the gain matrix $K_i$ has been obtained, the Jacobian of each phase-$i$, $A_i^d$, would be designed as

$$A_i^d = A_i - F_i K_i, 1 \leq i \leq N. \tag{22}$$

By the conventional DLQR method, the spectral radius of Jacobian $A_i^d$ would be smaller than 1. However, since the Jacobians $A_i^d$ designed by the DLQR method are generally not symmetric, we can then know from Remark 1 that the inequality condition $\rho(A_N^d ... A_2^d A_1^d) < 1$ may not be met, even though each Jacobian $A_i^d$ has spectral radius smaller than 1.

From Theorem 4, we know that the spectral radius of $\rho(A_N^d ... A_2^d A_1^d)$ would be smaller than 1 if the largest element of $A_i^d$ is smaller than $1/n$. Fortunately, this can be done by tuning the weighting matrix $Q_i$, which is associated with the robustness of the system. Increasing the $Q_i$ matrix leads to a smaller $A_i^d$. Therefore, we can choose a large enough $Q_i$ so that the largest element of $A_i^d$ is smaller than $1/n$.

Compared with the symmetric matrix method as well as the scale factor method, the DLQR-based method provides us with optimal feedback gain matrices in terms of the trade off between the robusness and the input energy. Hence this method may be more energy efficient. But on the other hand, the DLQR-based method would take manual efforts in picking appropiate weighting matrices $Q_i$.

## IV. Application to underactuated 3D bipedal locomotion

The objective of this section is to illustrate the continuous piecewise feedback control strategy to efficiently achieve the exponential stability of periodic 3D bipedal walking with multiple continuous phases. To show the result in a simple and clear way, a two-phase hybrid system is studied in this section. In addtion, to show that the proposed strategy is not limited to bipedal robots with left-right symmetry property which is assumed in [5, 16, 17], an underactuated 3D bipedal robot with asymmetric walking gait is considered in this section.

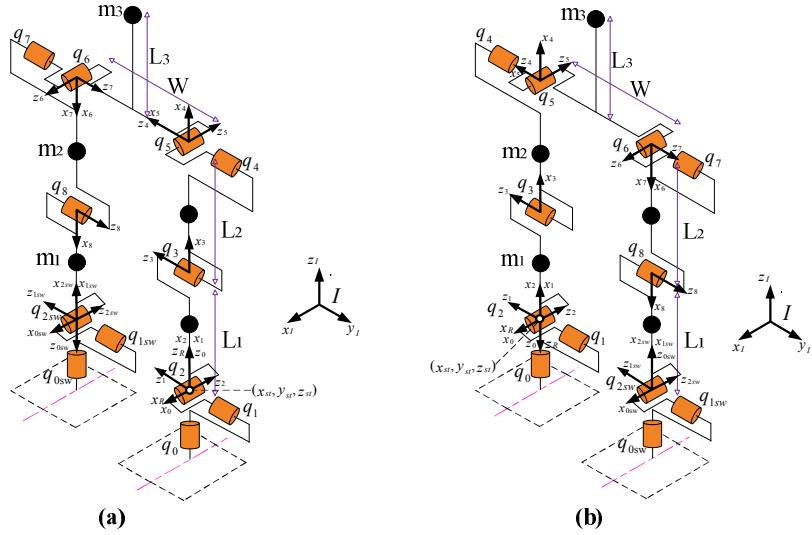

Fig. 2. The model and coordinate systems of the 3D five-link biped with point feet. In (a) and (b), the robot is supporting on the left and right legs, respectively. The structural parameters of the biped are chosen to be the same as those in [5]. The coordinate systems are established by the DH method. The virtual feet represented by the dashed box are used to show the yaw orientation of the biped.

*A. Hybrid models*

The 3D biped studied here is assumed to be underactuated with point feet, consisting of five links: a torso and two legs with knees, as shown in Fig.2. For simplicity, each link is modeled by a point mass at its center, and the stance leg end is assumed to act as a passive pivot. There are 9 DOF in total for the biped in the single support phase, i.e. $(q_0, q_1, .., q_8)$, including three underacuated DOF, i.e. $(q_0, q_1, q_2)$. In particular, the yaw angle, $q_0$, is assumed to be inhibited by friction during the whole single support phase. Therefore, the configuration coordinates of the robot in single support can be denoted simply by $q = (q_1, .., q_8)$.

The periodic walking gait studied in this section consists of two continuous phases, i.e., the two single support phases with support on the left and right legs, respectively. Let leg 1 be the left leg, and let leg 2 be the right leg. Without loss of generality, assume leg 1 is now the support leg. Based on the Lagrangian method, the dynamic model for the robot in the single support phase can be obtained as

$$D_1(q)\ddot{q} + C_1(q,\dot{q})\dot{q} + G_1(q) = Bu_1, \qquad (22)$$

where $D_1(q)$ is the positive-definite $(8\times 8)$ mass-inertia matrix, $C_1(q,\dot{q})$ is the $(8\times 8)$ Coriolis matrix, $G_1(q)$ is the $(8\times 1)$ gravity vector, $B$ is an $(8\times 6)$ full-rank, constant matrix indicating whether a joint is actuated or not, and $u_1$ is the $(6\times 1)$ vector of the input torques.

Under assumptions analogous to those in [5], the impact model in the double support phase can be obtained as

$$\begin{bmatrix} \dot{q}_e^+ \\ F_2 \end{bmatrix} = \begin{bmatrix} D_{1,e} & -E_{1,2}^T \\ E_{1,2} & 0_{4\times 4} \end{bmatrix}^{-1} \begin{bmatrix} D_{1,e}\dot{q}_e^- \\ 0_{4\times 4} \end{bmatrix}, \qquad (23)$$

where $\dot{q}_e^-$ and $\dot{q}_e^+$ are the extended velocities before and after the impact instantaneously, respectively, $F_2$ is the impulsive reaction force on leg 2 at the contact point, $D_{1,e}$ is the extended mass–inertia matrix, and $E_{1,2} = \partial/\partial q_e [x_{1,sw}, y_{1,sw}, z_{1,sw}, q_{0sw}]'$ is the Jacobian for the position of the swing foot and its orientation.

The support-swing leg swapping and relabeling take place immediately after the impact of the swing foot. This leads to the exchange between the notations of the angles, i.e., $[q_3, q_4, q_5, q_6, q_7, q_8] \to [q_8, q_7, q_6, q_5, q_4, q_3]$ [5]. At the same time, the angles $(q_0, q_1, q_2)$ exchange the roles with $(q_{0sw}, q_{1sw}, q_{2sw})$. Combining the impact model (23) and the coordinate relabeling, the dynamical model for the double support can be written as

$$\begin{bmatrix} q^+ \\ \dot{q}^+ \end{bmatrix} = \begin{bmatrix} \Delta_{1,q}(q^-) \\ \Delta_{1,\dot{q}}(q^-, \dot{q}^-) \end{bmatrix}, \qquad (24)$$

where $(q^-, \dot{q}^-)$ is the final state of the current step, and $(q^+, \dot{q}^+)$ is the initial state of the next step. Hence the double support model acts as a reset map.

When leg 2 is the support leg, the dynamic models can be obtained in the same way. Based on the above equations, the two-domain hybrid model for the biped robot can be established as

$$\Sigma = \begin{cases} X = \{X_i\}_{i=1}^{2} : X_i \subset R^{16} \\ U = \{U_i\}_{i=1}^{2} : U_i \subset R^{6} \\ FG = \{f_i, g_i\}_{i=1}^{2} : \dot{x}_i = f_i(x_i) + g_i(x_i)u_i \\ S = \{S_i^{i+1}\}_{i=1}^{2} : \{x_i | z_{i,sw}(q) = 0, x_{i,sw}(q) > 0\} \\ \Delta = \{\Delta_i^{i+1}\}_{i=1}^{2} : x_{i+1}^{+} = \Delta_i^{i+1}(x_i^{-}) \end{cases}, \quad (25)$$

where

$$f_i(x) = \begin{bmatrix} \dot{q} \\ -D_i^{-1}(C_i \dot{q} + G_i) \end{bmatrix}, \ g_1(x) = \begin{bmatrix} 0 \\ D_i^{-1} B \end{bmatrix},$$
$$x_{i+1}^{+} = \Delta_i(x_i^{-}) = \begin{bmatrix} \Delta_{i,q}(q^{-}) \\ \Delta_{i,\dot{q}}(q^{-}, \dot{q}^{-}) \end{bmatrix} \quad (26)$$

*B. Asymmetric walking gait design and stability analysis*

In order to implement the control strategy of this paper on a robot, a periodic and asymmetric walking gait was obtained as shown in Fig.3. Analogous to [5], the design of the gait was cast as a constrained nonlinear optimization problem, and the virtual constraints approach was also used here to design the output of the control for each phase-*i*, i.e.,

$$y_i = h_i(q) = q_a - h_{i,d}(\theta), i = 1, 2. \quad (27)$$

where $y_i$ denotes the output of control design, $q_a = [q_3, q_4, q_5, q_6, q_7, q_8]'$ is the vector of the actuated coordinates, $\theta = \theta(q)$ is a quantity strictly monotonically increasing along the walking cycle, and $h_{i,d}(\theta)$ is the desired evolution of the actuated variables as a function of $\theta$. Specifically, $h_{1,d}(\theta)$ and $h_{2,d}(\theta)$ are designed as the Bezier polynomials with 5 and 3 degrees, respectively. The whole optimization process was performed in MATLAB with the *Patternsearch* function of the optimization toolbox.

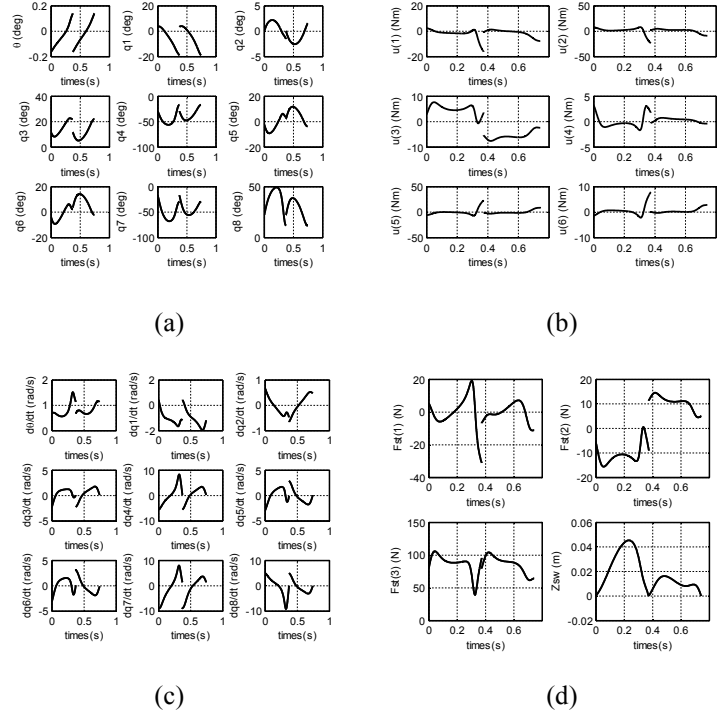

Fig. 3. The periodic and asymmetric walking motion of the 3D biped over two steps, where (a) is the joint angular profiles of the periodic motion, (b) is the required torque profiles of the periodic motion, (c) is the joint angular velocity profiles of the periodic motion, and (d) is the reaction force on the stance foot and the evolution of the swing leg tip.

To save the computational cost, the numerical integration of the single support phase dynamics was carried out on a lower-dimensional system, which is called the swing phase zero dynamics, see [5]. It is noted that, during the optimization, the walking gait is feasible only when the reaction forces $F_{st}$ on the support leg meets the no-take-off and friction constraints. The calculation of $F_{st}$ is given in Appendix B.

For a single-domain hybrid system, the exponential stability of the full hybrid model can be evaluated by the corresponding reduced hybrid zero dynamics. According to Theorem 1, we can also apply this method to the two-domain hybrid model in this section. Firstly, a zero dynamics of the full hybrid model has to be created. To achieve that, a correction term $h_{i,c}$ was introduced to the output in each phase-$i$.

$$y_i = h_i(q) = q_a - h_{i,d}(\theta) - h_{i,c}, i = 1, 2. \tag{28}$$

where $h_{i,c}$ was designed so that the initial errors of the outputs and their derivatives,

which are caused by the impact event, can be smoothly joined to the original virtual constraint at the middle of the step, see [5].

After the HZD for the two-domain hybrid model had been created, two restricted partial maps were then constructed, i.e., $P_1: S_1^2 \cap Z \to S_2^1 \cap Z$, $P_2: S_2^1 \cap Z \to S_1^2 \cap Z$ where $Z = \{(q, \dot{q}) \mid \dot{y}_i(q) = 0, \dot{y}_i(q) = 0\}$. Following the technique in [5], the Jacobians for $P_1$ and $P_2$ can be calculated as

$$A_1 = \begin{bmatrix} 2.0824 & -0.3380 & 0.2596 \\ 2.2540 & -1.2701 & -4.5384 \\ 7.1221 & -2.8911 & -1.1735 \end{bmatrix} \quad (29)$$

$$A_2 = \begin{bmatrix} 2.1280 & 0.2878 & 0.3763 \\ -2.2854 & -1.3484 & 4.7298 \\ 7.3238 & 2.6283 & -0.5894 \end{bmatrix} \quad (30)$$

respectively. According to Theorem 2, the Jacobian of the Poincaré return map $P$ can be obtained as

$$A = \begin{bmatrix} 7.7603 & -2.1727 & -1.1953 \\ 25.8883 & -11.1895 & -0.0243 \\ 16.9777 & -4.1095 & -9.3347 \end{bmatrix} \quad (31)$$

with the three eigenvalues:

$$\begin{aligned} \lambda_1 &= 2.8271, \quad \lambda_{2,3} = -7.7955 \pm 2.2193i \\ |\lambda_{2,3}| &= 8.1053 \end{aligned} \quad (32)$$

Since $\rho(A) = 8.1053 > 1$, the designed asymmetric gait is unstable. Therefore, we need to design a feedback controller to stabilize the hybrid system.

### C. Achieving exponential stability by the continuous piecewise feedback control strategy

The objective of this subsection is to stabilize the unstable asymmetric walking gait by the continuous piecewise feedback control strategy. Firstly, a continuous time controller $\Gamma_i(x_i, \beta_i)$ is designed for each continuous phase. Secondly, the scale factor method is applied to design the feedback parameters to stabilize the hybrid system.

In the first step, a continuous time controller $\Gamma_i(x_i,\beta_i)$ is designed for each continuous phase based on the result of Lemma 1. From the result in [13], we know that the nominal input torque $u_i$ is determined by the output $y_i$, and thus $u_i$ can be expressed as $u_i(y_i)$. Therefore, from Lemma 1, $\Gamma_i(x_i,\beta_i)$ can be designed as

$$\Gamma_i(x_i,\beta_i) = u_i(y_i + \Omega(x_i,\beta_i)), i=1,2, \qquad (33)$$

where $\Omega(x_i,\beta_i)$ satifies the conditions $\lim_{\beta_i \to 0}\Omega(x_i,\beta_i)=0$, and $\lim_{\beta_i \to 0}\partial\Omega(x_i,\beta_i)/\partial x_i = 0$. Specifically, $\Omega(x_i,\beta_i)$ is designed in the form of $\Omega(\theta,\beta_i)$. It follows that the ouput (28) will be updated as

$$y_i = h_i(q) = q_a - h_{i,d}(\theta) - h_{i,c} + \Omega(\theta,\beta_i), i=1,2. \qquad (34)$$

It is noted that one of the other important requirements for the design of $\Omega(\theta,\beta_i)$ is to preserve the HZD that have been created in the above section. Here, $\Omega(\theta,\beta_i)$ is designed as a fifth-order polynomial.

In the second step, the scale factor method is applied to design the feedback parameters to stabilize the hybrid system. Firstly, Similar to the calculation of $A_1$ and $A_2$, the Jacobians of $P_1$ and $P_2$ with respect to $\beta_1$ and $\beta_2$ are obtained as

$$F_1 = \begin{bmatrix} -0.0326 & -0.0369 & -0.1358 & 0.0604 & 0.0396 & 0.0229 \\ 0.0890 & -0.1037 & -0.0422 & 0.0306 & -0.4580 & -0.1079 \\ -0.1596 & -0.2808 & -0.3050 & 0.6396 & 0.3171 & 0.1466 \end{bmatrix}, \qquad (35)$$

$$F_2 = \begin{bmatrix} 0.0483 & 0.0397 & -0.1208 & 0.0589 & -0.0445 & -0.0259 \\ 0.0868 & -0.0880 & 0.0529 & -0.0243 & -0.4689 & -0.1057 \\ 0.2465 & 0.2798 & -0.2130 & 0.6212 & -0.3389 & -0.1624 \end{bmatrix}, \qquad (36)$$

Let $m_1 = \max_{1 \le j,k \le n} |A_1(j,k)|$, $m_2 = \max_{1 \le j,k \le n} |A_2(j,k)|$. Then, the scale factors for the two partial maps can be designed as $c_1 = 1/(n.m_1)$ and $c_2 = 1/(n.m_2)$. From (20), we have

$$K_1 = \begin{bmatrix} -1.4696 & -0.0063 & -1.5780 \\ -3.8313 & 2.2786 & 3.9165 \\ -12.1720 & 0.5296 & -2.5886 \\ 4.0627 & -3.9668 & -5.5839 \\ -2.8061 & 1.7885 & 7.7203 \\ 0.3861 & 0.1001 & 1.6090 \end{bmatrix}, \quad (37)$$

$$K_2 = \begin{bmatrix} 3.1276 & 0.3810 & 1.3404 \\ 4.6644 & 2.2536 & -3.5804 \\ -12.4826 & -0.2466 & -3.0383 \\ 4.9906 & 3.9753 & -5.8621 \\ 2.8635 & 2.1413 & -8.3819 \\ -0.7739 & 0.0753 & -1.6155 \end{bmatrix}, \quad (38)$$

And the corresponding Jacobians are designed as

$$A_1^d = \begin{bmatrix} 0.0975 & -0.0158 & 0.0122 \\ 0.1055 & -0.0594 & -0.2124 \\ 0.3333 & -0.1353 & -0.0549 \end{bmatrix}, \quad (39)$$

$$A_2^d = \begin{bmatrix} 0.0969 & 0.0131 & 0.0171 \\ -0.1040 & -0.0614 & 0.2153 \\ 0.3333 & 0.1196 & -0.0268 \end{bmatrix}, \quad (40)$$

Finally, according to (14), the Jacobian of the Poincaré return map $P$ are designed as

$$A^d = \begin{bmatrix} 0.0166 & -0.0046 & -0.0025 \\ 0.0551 & -0.0238 & 0 \\ 0.0363 & -0.0088 & -0.0199 \end{bmatrix}, \quad (41)$$

which has a spectral radius of 0.0173. Hence, the closed-loop system is exponential stable.

To illustrate the effect of the continuous piecewise feedback control strategy, the HZD model of the 3D biped in a closed-loop is simulated with the initial state perturbed away from the fixed point $x_1^*$, and the initial errors of 0.01, 0.01 and 0.01 are introduced to the three components of $x_1^*$, respectively. Fig. 4 shows the phase plots of the uncontrolled variables $\theta$ and $q_2$, from which the asymmetric walking gait is apparently stabilized by the continuous piecewise feedback control strategy.

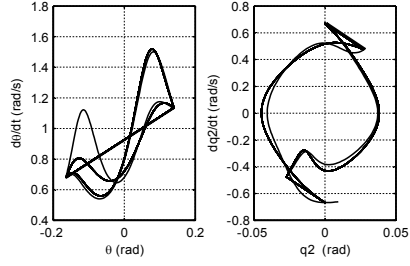

Fig. 4. Phase plots for $\theta$ and $q_2$, where the straight lines correspond to two impact phases in one walking cycle.

## V. Conclusion

This paper addresses the problem of exponentially stabilizing the periodic orbits of multi-domain hybrid systems arising from 3D bipedal walking. By exploiting the properties of the Poincaré maps regarding the multi-domain hybrid systems, a continuous piecewise feedback control strategy is presented, and three methods are furthermore given to design controller parameters based on the developed theorems. In contrast with previous methods that require the controller parameters in each continuous phase to be designed simultaneously, the proposed method is able to design the controller parameters in each continuous phase independently, which allows the strategy to be applied to hybrid systems with multiple domains. The validity of this approach was illustrated in an underactuated 3D bipedal robot with asymmetric walking gait.

The strategy presented in this paper can be extended to more general forms of 3D bipedal locomotions, including running and human-like walking. In future research, we will investigate the robustness of this strategy against all sources of uncertainties in 3D bipedal walking as we have done in the previous works [19, 20].

## Appendix

***A. Approximation of*** $A_i^d = A_{i,\beta} - F_{i,\beta} K_i$

Let $f_{op} = f_i(x_i) + g_i(x_i) u_i(x_i)$, and let $f_{cl} = f_i(x_i) + g_i(x_i) \Gamma_i(x_i, \beta_i)$. Then, under the conditions C1 and C2, we can easily prove that

$$\lim_{\beta_i \to 0} f_{op} = f_{cl}, \tag{42}$$

And

$$\lim_{\beta_i \to 0} \partial f_{cl}(x_i, \beta_i) / \partial x_i = \partial f_{op}(x_i, \beta_i) / \partial x_i. \tag{43}$$

From (42) and (43), we can know that, when the control parameter vector $\beta_i$ is close to zero, the nominal trajectory generated by $\dot{x}_i = f_{cl}$ equals that of $\dot{x}_i = f_{op}$. In other words, the periodic orbit is invariant when $\beta_i$ is close to zero. Combining the result in [17], where the mathematical expression of $\partial P_{i,\beta} / \partial x_{i-1}^S$ is given, we will have

$$\lim_{\beta_i \to 0} \partial P_{i,\beta} / \partial x_{i-1}^S = \partial P_i / \partial x_{i-1}^S. \tag{44}$$

Therefore, for a small enough vector $\beta_i$, we think that $\partial P_i / \partial x_{i-1}^S$ is approximately equal to $\partial P_{i,\beta} / \partial x_{i-1}^S$, i.e., $A_{i,\beta} = A_i$, where $A_i = \partial P_i / \partial x_i^S(x_i^*)$.

When it comes to the estimation of $F_{i,\beta}$, it is much easier. Since $P_{i,\beta}(x_{i-1}^S, \beta_i) := \phi_i(T_{I,i}(\Delta_{i-1}^i(x_{i-1}^S)), \Delta_{i-1}^i(x_{i-1}^S), \beta_i)$ is $C^{n_p}$ with respect to $\beta_i$, we can then know that $F_{i,\beta} = \partial P_{i,\beta} / \partial \beta_i$ is continuous. Hence, we have

$$\lim_{\beta_i \to 0} F_{i,\beta}(\beta_i) = F_{i,\beta}(0). \tag{45}$$

For convenience, $F_{i,\beta}(0)$ will be denoted as $F_i$. Replacing (44) and (45) in (10), we have

$$A_i^d = A_i - F_i K_i, \tag{46}$$

where $A_i^d$ is independent of $\beta_i$, $A_i$ and $F_i$ can be calculated by following the techniqie in [5].

### B. Proof of Theorem 3

In the first step of the proof, we will prove that, for any symmetric matrix $A \in R^{n \times n}$, its spectral norm is equivelent to its spectral radius, i.e. $\|A\|_2 = \rho(A)$.

From the definition of spectral norm, we have

$$\|A\|_2 = \sqrt{\max_i |\lambda_i((A)^H A)|} = \sqrt{\rho(A^H A)}. \tag{47}$$

Since $A$ is symmetric, the above equation can be rewritten as

$$\|A\|_2 = \sqrt{\rho(A^H A)} = \sqrt{\rho(A^2)}. \tag{48}$$

From the matrix theory, we can know that if $\lambda_1, \lambda_2, \ldots, \lambda_s$ are the eigenvalues of $A$, then $(\lambda_1)^k, (\lambda_2)^k, \ldots, (\lambda_s)^k$ would be the eigenvalues of $A^k$. Hence, we have

$$\rho(A^k) = \max_i |(\lambda_i)^k| = \left(\max_i |\lambda_i|\right)^k = [\rho(A)]^k. \tag{49}$$

Combining (48) and (49), we have

$$\|A\|_2 = \sqrt{\rho(A^2)} = \sqrt{\rho^2(A)} = \rho(A). \tag{50}$$

The next step is to prove that, for two designed symmetric Jabocians, such as $A_j^d$ and $A_k^d$, where $1 \leq j, k \leq n$, if $\rho(A_j^d) < 1$ as well as $\rho(A_k^d) < 1$, then $\rho(A_j^d A_k^d) < 1$.

According to Theorem 5.6.9 in [21], we have

$$\rho(A_j^d A_k^d) \leq \|A_j^d A_k^d\|_2 \leq \|A_j^d\|_2 \|A_k^d\|_2. \tag{51}$$

Since $A_j^d$ and $A_k^d$ are symmetrix matrixes, from (50), we have $\rho(A_j^d) = \|A_j^d\|_2$, $\rho(A_k^d) = \|A_k^d\|_2$. Hence, (51) can be rewritten as

$$\rho(A_j^d A_k^d) \leq \|A_j^d\|_2 \|A_k^d\|_2 = \rho(A_j^d)\rho(A_k^d) < 1, \tag{52}$$

Based on (52), Theorem 3 can be proved in a further step.

## *C. Proof of Theorem 4*

Firstly, we will prove that, for two designed Jabocians, such as $A_j^d$ and $A_k^d$, where $1 \leq j, k \leq n$, if $m_j = \max_{1 \leq i, k \leq n} |A_j^d(i,k)| \leq 1/n$ as well as $m_k = \max_{1 \leq i, j \leq n} |A_k^d(i,j)| \leq 1/n$, then $\rho(A_j^d A_k^d) < 1$.

From matrix theory, for any $A \in R^{n \times n}$, define a norm $\|A\|_\infty = \max_{1 \leq i, j \leq n} |a_{ij}|$, and then $N(A) = n\|A\|_\infty$ is a matrix norm. According to Theorem 5.6.9 in [21], we can then have

$$\rho(A_j^d A_k^d) \leq N(A_j^d A_k^d). \tag{53}$$

Let $A_{com} = A_j^d A_k^d$. Then, the $(l_1, l_2)$ entry of $A_{com}$ can be expressed as $A_{com}(l_1, l_2) = \sum_{m=1}^{n} A_j^d(l_1, m) A_k^d(m, l_2)$, where $1 \leq l_1, l_2 \leq n$. Since that $m_j = \max_{1 \leq i,k \leq n} |A_j^d(i,k)| \leq 1/n$, $m_k = \max_{1 \leq i,j \leq n} |A_k^d(i,j)| \leq 1/n$, we can then easily prove that, $|A_{com}(l_1, l_2)| < 1/n$, for any $l_1, l_2$, i.e.,

$$\|A_{com}\|_\infty = \max_{1 \leq l_1, l_2 \leq n} |A_{com}(l_1, l_2)| < 1/n. \tag{54}$$

Hence $N(A_{com}) = n \cdot \|A_{com}\|_\infty < 1$. Therefore, (53) can be rewritten as

$$\rho(A_j^d A_k^d) \leq N(A_j^d A_k^d) = N(A_{com}) < 1. \tag{55}$$

Based on (55), Theorem 4 can be proved in a further step.

### D. Calculation of the ground reaction forces on the stance leg

Suppose the biped is supporting on leg 1. Then, based on the Lagrangian method, the dynamical model of the single support phase is developed by the full set of generalized coordinates $q_e = [x_{st}, y_{st}, z_{st}, q_0, q_1, ..., q_8]'$, i.e.

$$D_{1,e}(q_e)\ddot{q}_e + C_{1,e}(q_e, \dot{q}_e)\dot{q}_e + G_{1,e}(q_e) = \begin{bmatrix} 0_{4 \times 1} \\ Bu_1 \end{bmatrix} + \begin{bmatrix} F_{st} \\ 0_{8 \times 1} \end{bmatrix}, \tag{56}$$

where $D_{1,e}(q_e)$ is the positive-definite $(12 \times 12)$ mass-inertia matrix, $C_{1,e}(q_e, \dot{q}_e)$ is the $(12 \times 12)$ Coriolis matrix, $G_{1,e}(q_e)$ is the $(12 \times 1)$ gravity vector, and $F_{st}$ is the $(4 \times 1)$ ground reaction force vector on the stance leg. When we partition the matrices $D_{1,e}(q_e)$, $C_{1,e}(q_e, \dot{q}_e)$, $G_{1,e}(q_e)$ into the corresponding submatrices, (56) can be rewritten by

$$\begin{bmatrix} D_{1,e11} & D_{1,e12} \\ D_{1,e21} & D_{1,e22} \end{bmatrix} \begin{bmatrix} \ddot{q}_{pos} \\ \ddot{q} \end{bmatrix} + \begin{bmatrix} C_{1,e11} & C_{1,e12} \\ C_{1,e21} & C_{1,e22} \end{bmatrix} \begin{bmatrix} \dot{q}_{pos} \\ \dot{q} \end{bmatrix} + \begin{bmatrix} G_{1,e11} \\ G_{1,e21} \end{bmatrix} = \begin{bmatrix} 0_{4 \times 1} \\ Bu_1 \end{bmatrix} + \begin{bmatrix} F_{st} \\ 0_{8 \times 1} \end{bmatrix}, \tag{57}$$

where $\ddot{q}_{pos} = [x_{st}, y_{st}, z_{st}, q_0]'$, $q = [q_1,...,q_8]'$. From the first row of (57), we obtain

$$D_{1,e11}\ddot{q}_{pos} + D_{1,e12}\ddot{q} + C_{1,e11}\dot{q}_{pos} + C_{1,e12}q + G_{1,e11} = F_{st}. \tag{58}$$

Since the values of the components of $\ddot{q}_{pos} = [x_{st}, y_{st}, z_{st}, q_0]'$ are always zero in the single support phase, the ground reactive forces $F_{st}$ can be calculated from (58) and we have

$$F_{st} = D_{1,e12}\ddot{q} + C_{1,e12}q + G_{1,e11}. \tag{59}$$